# A Figurative Identification for Superposed OAM Modes in FSO Systems

**Haowei Shi[1,2], Mutong Xie[1], Xinlu Gao[1,2], Shanguo Huang[1]**

*1. State Key Laboratory of Information Photonics and Optical Communications, Beijing University of Posts and Telecommunications*

*2. School of Science, Beijing University of Posts and Telecommunications*

*Beijing 100876, China, xinlugao@bupt.edu.cn*

**Abstract:** We demonstrate that a complete projection in Hilbert Space figuratively describes a superposed state, introducing a new scale to qualify an FSO system. Measurement simulation of superposed OAM beam through this projection scheme is given.

## 1. Introduction

The orbital angular momentum (OAM) modes have been long exploited to develop a spatial multiplexed free space optical (FSO) communication system [1]. The generation of OAM has been realized with a variety of apparatus, e.g. spiral phase plate, q-plate, plasmonic metasurfaces, forked gratings [2], etc. However, its effectiveness has been assailed that the channel capacity of OAM modes is limited, under the potential maximum [3]. Fortunately, the promising advantage of OAM modes is not confined to the enhancement of channel capacity. An OAM-coded signal has been proved to possess the exclusive anti-turbulence property when it comes to the strong atmosphere turbulence [4]. Meanwhile, the existence of OAM-carried waves has been demonstrated in rotational Doppler effect [5]. All of the above call attention to the identification of superposed OAM modes, which has been neglected in communication system but pervasive in the nature. In this thesis a geometric-like identification of OAM modes is illustrated based on its description in Hilbert space. A method for the quantitative measurement is suggested for reference.

## 2. Description of OAM modes in Hilbert Space

According to Gleason's [6], a square-integrable function can be completely projected to a set of Fourier basis. The basis functions can be viewed as basis vectors in Hilbert Space. Hence any superposed OAM mode can be projected, and viewed as a composite vector composed of those which are parallel to coordinate axes as is shown in Fig. 1. When it comes to OAM modes, the basis vectors are specifically Laguerre-Gaussian functions. Define $f_l(\varphi) = e^{il\varphi}, f_l g_p(r,\varphi) = L_p^l(r)e^{il\varphi}$. A given mode $[a_{11}L_1^1(r) + a_{13}L_3^1(r)]e^{i\varphi} + [a_{31}L_1^3(r) + a_{32}L_2^3(r)]e^{i3\varphi}$, marked with the black vector, can be measured experimentally in intensity respectively projected to the Hilbert basis vectors, where the projection is defined as:

$$< \vec{E}, fg_i(x,y) > = \iint \sum_j E_j fg_j(x,y) \cdot fg_i(x,y) dxdy = \sum_j E_j \delta(i-j) = E_i \qquad (1)$$

Please notice that the basis $f_l g_p(r,\varphi)$ presented in the figure, tailored to Laguerre-Gaussian functions, cannot be decomposed to $f_l(r)g_p(\varphi)$ since $l$ and $p$ are coupled.

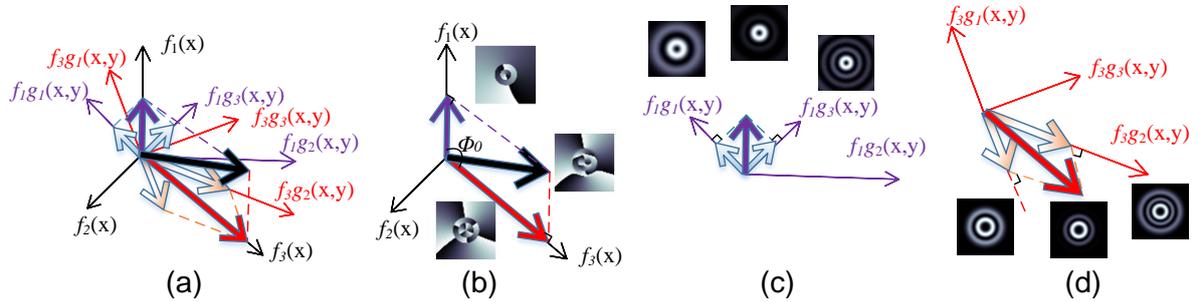

Fig. 1 An archetype of vector projection. (a) Black vectors are firstly projected to $V_1 \times V_2$ as the superposition of the purple vector and the red vector, then projected to $V_2$ respectively on corresponding basis, where $V_1(x) = \{\sum_i a_i \vec{f}_i \mid a_i \in R\}$, $V_2(y) = \{\sum_i b_i \vec{g}_i \mid b_i \in R\}$. (b) Projection to $V_1$ of the black, shown in phase pattern. (c), (d) Projection to $V_2$ respectively of the $f_1$, $f_3$ part, shown in intensity pattern.

Practically, the deviation of a tested signal from a specific mode is widely urged in communication systems and optical detection, which should better be independent with the magnitude of the signal. Here a new scale of deviation is illustrated as an analogy of that in a Euclidean space. In a geometric space, the difference between two vectors $\vec{a}$ and $\vec{b}$ is scaled with the angle $\varphi$ between them, which is defined as $\varphi = \arccos(<\vec{a},\vec{b}>) = \arccos(\sum_i a_i b_i)$.

In Hilbert space, between $\vec{E_1}$ and $\vec{E_2}$ we define Hilbert angle $\Phi = \arccos(<\vec{E_1},\vec{E_2}>)$, where

$$<\vec{E_1},\vec{E_2}> = \sum_{i,j} E_{1i} E_{2j} \iint f_i(x,y) f_j(x,y) dxdy = \sum_{i,j} E_{1i} E_{2j} \delta(i-j) = \sum_i E_{1i} E_{2i} \qquad (2)$$

Specifically, the marked Hilbert angle $\Phi_0$ is exactly $\pi/2$, due to the orthogonality of $\{f_1(x), f_3(x)\}$, or quantitatively calculated as

$$\cos\Phi_0 = \iint A f_1(x) g_A(y) B f_3(x) g_B(y) dxdy = AB \int g_A(y) g_B(y) dy \int f_1(x) f_3(x) dx = 0 \qquad (3)$$

The properties of angle in Euclidean space are fully transferred to that in Hilbert space. $\Phi$ is qualified with **1. one-to-one correspondence with the deviation**, 2. **Magnitude-Independence**, 3. **Normality**. Hence specifically the Hilbert angle scales the energy efficiency of the given communication system.

## 3. Experiment Result

Conventionally, the measurement of a signal vector is operated as a projection to Hilbert space. As to OAM basis, the components of each OAM modes are measured with a 4f system modified with an SLM modulated as an M-order forked grating. The experiment setup is shown in Fig. 2 (a).

According to angular spectrum method, the propagation of a Bessel beam through such a modified 4f system can be simplified as an transfer operator $\hat{C}$, where

$$C_{mn} = <J_m e^{im\varphi} | \hat{C} | J_n e^{in\varphi}> = \delta(m - n + M) \qquad (4)$$

The clarity results from the equivalence of the radial spectrum of all Fourier-Bessel basis functions despite the different azimuthal orders, which reduces the dispersion induced by SLM to a minimum [7]. Noticing that the output Bessel modes never converges except when the azimuthal order m=0, only in order n = M is the input Bessel mode focused on the center of object screen while the others forms nonzero-order Bessel rings. Meanwhile, Bessel modes form a complete set of field function with finite energy, including Laguerre modes. Hence all OAM beams of M order, namely an azimuthal factor $e^{im\varphi}$, converge at the center spot isolated from those of non-M order. Taken as an example, a measurement simulation of an OAM beam of order 3 superposed on 6 is illustrated in Fig. 2 (c, d, e, f). A matrix of detectors based on M = 3, 6, 9, 12 forked grating is illustrated by measuring the magnitude at the specific points on the output screen.

Furthermore, the theory, applied to the analysis of the spectrum from an antenna array, provides an intrinsic outlook of the relationship between $\Phi$ the energy efficiency of communication and N the quantity of antennas. The result of simulation is illustrated in Fig. 3 and Tab. 1. The intricate fact is totally manifested by the scale $\Phi$ that the discrete sampling dissipates the energy from the initial frequency mode to the whole frequency domain. The more frequently sampled it is, the smaller is the deviation $\Phi$.

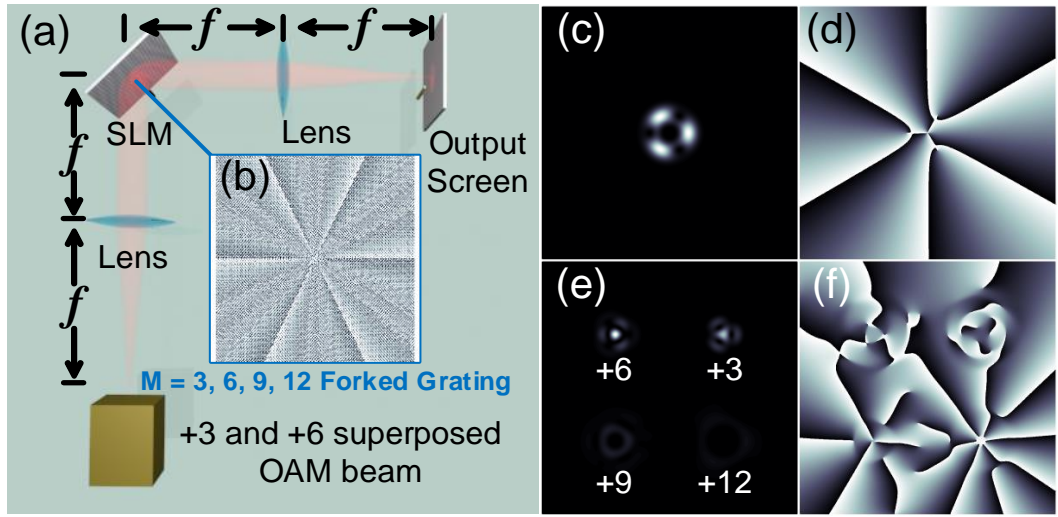

Fig. 2 (a) The setup of the OAM identification experiment. (b) The forked grating with M = 3, 6, 9, 12 mounted on to the SLM. (c) Input intensity and (d) input phase of the superposed OAM beam including azimuthal order +3 & +6. (e) Output intensity after the SLM. Spots representing state +3 and +6 are noticeable while those of +9 and +12 are negligibly faint. (f) Output phase after the SLM.

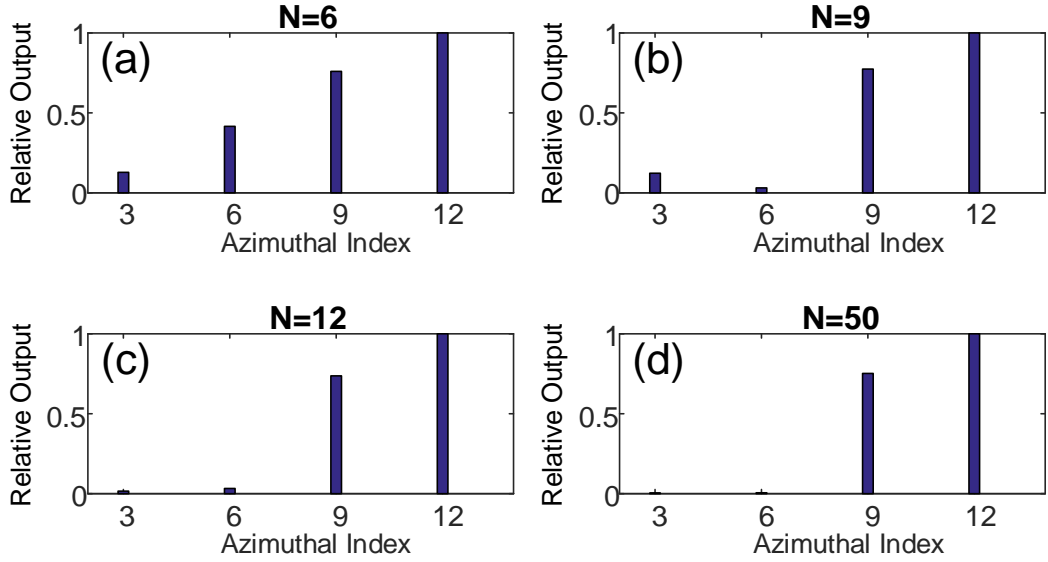

Fig. 3 The output spectrum $\Phi$ from discrete antenna arrays. The input vector is set with the azimuthal indices +9, +12. The expected output vector is (0,0,1,1.45), where the input (0,0,1,1) is partially received and the energy loss varies with respect to the different azimuthal order.

Tab. 1. The Hilbert angle $\Phi$ from discrete antenna arrays.

| N | 6 | 9 | 12 | 50 |
|---|---|---|----|----|
| $\Phi$/rad | 0.3504 | 0.1257 | 0.0754 | 0.0123 |

## 4. Summary

Tailored to the increasingly diverse demand, a novel perspective of OAM modes is illustrated in this thesis. In geometric analogy, the normalized scale of angular deviation well describes the extent to which the pattern deviates from OAM modes. A quantitative optimization method for projection measurement in FSO systems is offered based on a specifically modified $4f$ system. Furthermore, the methodology of projection is demonstrated in Hilbert space, which means a potential to be extended into more universal finite signal analysis.

## 5. Acknowledgements

This work is supported in part by the National Natural Science Foundation of China (NSFC) (61690195, 61575028, 61605015); the National Science Foundation for Outstanding Youth Scholars of China (61622102); the Fundamental Research Funds for the Central Universities (2016RCGD21, 2016RC26); the Open Funds of IPOC (IPOC2016ZT03) and UWC (KFKT-2015102).